\documentclass[pra,superscriptaddress,twocolumn,showpacs]{revtex4}
\usepackage{graphicx}
\usepackage{dcolumn}
\usepackage{amsmath}
\usepackage{amssymb}
\newcommand{\bfr}{\mathbf{r}}

\begin{document}

\title{Exchange and correlation in open systems of fluctuating
electron number}

\author{John P. Perdew}
\affiliation{Department of Physics, Tulane University, New Orleans,
Louisiana 70118, USA}
\author{Adrienn Ruzsinszky}
\affiliation{Department of Physics, Tulane University,
New Orleans, Louisiana 70118, USA}
\author{G\'abor I. Csonka}
\affiliation{Department of Chemistry, Budapest University
of Technology and Economics, H-1521 Budapest, Hungary}
\author{Oleg A.~Vydrov}
\affiliation{Department of Chemistry, Rice University,
Houston, Texas 77005, USA}
\author{Gustavo E. Scuseria}
\affiliation{Department of Chemistry, Rice University,
Houston, Texas 77005, USA}
\author{Viktor N. Staroverov}
\affiliation{Department of Chemistry, University of Western Ontario,
London, Ontario N6A 5B7, Canada}
\author{Jianmin Tao}
\affiliation{Department of Physics, University of Missouri--Columbia,
Columbia, Missouri 65211, USA}

\date{\today}

\begin{abstract}
While the exact total energy of a separated open system varies
linearly as a function of average electron number between
adjacent integers, the energy predicted by semi-local density
functional approximations curves upward and the exact-exchange-only
or Hartree-Fock energy downward. As a result, semi-local density
functionals fail for separated open systems of fluctuating electron
number, as in stretched molecular ions A$_2^{+}$ and in solid
transition metal oxides. We develop an exact-exchange theory and
an exchange-hole sum rule that explain these failures and we propose
a way to correct them via a local hybrid functional.
\end{abstract}

\maketitle

Kohn-Sham density functional
theory~\cite{Kohn:1965/PR/A1133,Book/Fiolhais:2003} (DFT)
replaces the correlated wavefunction problem by a
more tractable problem of non-interacting electrons moving
in self-consistent effective potentials $v_s^\sigma(\bfr)$
($\sigma=\uparrow,\downarrow$) which generate the spin densities
$n_{\sigma}(\bfr)$ of the real (interacting) system. Exact in principle
for the ground-state energy and density, Kohn-Sham DFT requires
in practice an approximation for the exchange-correlation (xc)
energy functional $E_\mathrm{xc}[n_\uparrow,n_\downarrow]$.
With improving approximations, DFT has become the standard method
for electronic structure calculations in physics and chemistry.

In terms of the total electron density $n=n_\uparrow+n_\downarrow$
and exchange-correlation energy per electron
$\epsilon_\mathrm{xc}(\bfr)$, we write
\begin{equation}
 \vspace*{-0.20cm}
 E_\mathrm{xc} = \int d\bfr\,n(\bfr) \epsilon_\mathrm{xc}(\bfr).
   \label{exc}
\end{equation}
A ladder~\cite{Perdew:2001/1} of approximations constructs
$\epsilon_\mathrm{xc}(\bfr)$ as a function of density-dependent
ingredients. The first three rungs are semi-local
(with $\epsilon_\mathrm{xc}$ found from the Kohn-Sham orbitals
in an infinitesimal neighborhood of $\bfr$) and in some versions
non-empirical. The rungs are defined by the ingredients: (i) the
local spin-density (LSD) approximation~\cite{Kohn:1965/PR/A1133},
which uses only $n_\sigma(\bfr)$; (ii) the generalized gradient
approximation (GGA) in the Perdew-Burke-Ernzerhof
(PBE)~\cite{Perdew:1996/PRL/3865} version, which adds the gradients
$\nabla n_\sigma(\bfr)$; (iii) the meta-GGA in the
Tao-Perdew-Staroverov-Scuseria (TPSS)~\cite{Tao:2003/PRL/146401}
version, which further adds the positive orbital
kinetic energy densities $\tau_\sigma(\bfr)$;
(iv) functionals employing a truly nonlocal ingredient,
the exact (ex) exchange energy density $n\epsilon_\mathrm{x}^\mathrm{ex}$,
either in full~\cite{Perdew:2001/1,Mori-Sanchez:2006/JCP/091102}
or in part~\cite{Jaramillo:2003/JCP/1068,Becke:2005/JCP/064101}.
The currently used fourth-rung functionals are global
hybrids (gh)~\cite{Becke:1993/JCP/5648}, mixtures of exact exchange
and semi-local (sl) approximations
\begin{equation}
 \vspace*{-0.1cm}
 E_\mathrm{xc}^\mathrm{gh} = aE_\mathrm{x}^\mathrm{ex}
 + (1-a)E_\mathrm{x}^\mathrm{sl} + E_\mathrm{c}^\mathrm{sl},
  \label{global}
\end{equation}
where the exact-exchange mixing coefficient ``$a$'' is a global
empirical parameter (typically $a \approx 0.2$). A global hybrid
with $0 \le a < 1$ does not satisfy any universal constraints beyond
those satisfied by $E_\mathrm{xc}^\mathrm{sl}$.

In many real systems, these existing functionals are reasonably
accurate for $E_\mathrm{x}$ and more accurate (due to error
cancellation) for $E_\mathrm{xc}$, with accuracy generally
increasing up the ladder. Yet serious errors occur in
nearly-separated open systems with fluctuating electron numbers
that may not average to integer values, as summarized below.
(a) In the dissociation of heteronuclear diatomics such as NaCl
with bond length $R$, spurious fractional-charge $R\to\infty$
limits are common (e.g., Na$^{+0.4}\cdots$Cl$^{-0.4}$ instead of
Na$^{0}\cdots$Cl$^{0}$), as are related charge-transfer errors.
(In this example, Na$\cdots$Cl is a closed
system of fixed integer electron number, while Na and Cl are separated
open subsystems free to exchange electrons with each other).
(b) In the dissociation of molecular radical cations A$_2^{+}$,
the $R\to\infty$ limit is correctly A$^{+0.5}\cdots$A$^{+0.5}$,
but the total energy is far below that of A$\cdots$A$^{+}$, with
which it should be degenerate. For the one-electron molecule H$_2^{+}$,
this is unambiguously~\cite{Ruzsinszky:2005/JPCA/11006}
a self-interaction error. The errors (a) and (b) are not
necessarily corrected by functionals that are exact for all
one-electron densities~\cite{Ruzsinszky:2006-2007}.
(c) In the solid state, energy competition among electronic
configurations in transition metal oxides, lanthanides, and
actinides can be poorly described~\cite{Kudin:2002/PRL/266402}.
These errors are similar in LSD, PBE and TPSS, but are improved
by global hybrids~\cite{Kudin:2002/PRL/266402}.  In this Report,
we derive the generalized x-hole sum rule and then explain at the
most fundamental level why semi-local functionals fail for open
systems by showing that they violate this rule. We also show that
the errors of semi-local functionals for $E_\mathrm{xc}$ can be
corrected by a properly designed local hybrid functional.

For closed systems of integer electron number and integer occupation
numbers, the exact (Hartree-Fock-like) exchange energy per electron
at $\bfr$ is given by
\begin{equation}
\epsilon_\mathrm{x}(\bfr) = \frac{1}{2}\int d\bfr' \,
  \frac{n_\mathrm{x}(\bfr,\bfr')}{|\bfr - \bfr'|},
  \label{exden}
\end{equation}
where $n_\mathrm{x}(\bfr,\bfr')$ is the density at $\bfr'$ of
the exchange hole around an electron at $\bfr$:
\begin{equation}
 n_\mathrm{x}(\bfr,\bfr')
  = -\sum_{\sigma}\frac{|\rho_{\sigma}(\bfr,\bfr')|^2}{n(\bfr)},
  \label{xhole}
\end{equation}
in which
\begin{equation}
\rho_{\sigma}(\bfr,\bfr') = \sum_\alpha f_{\alpha\sigma}
 \psi_{\alpha\sigma}^{*}(\bfr') \psi_{\alpha\sigma}(\bfr)
 \label{matrix}
\end{equation} 
is the one-particle density matrix for spin $\sigma$ of the Kohn-Sham
non-interacting system, $\psi_{\alpha\sigma}$ are orbitals and
$f_{\alpha\sigma}$ their occupation numbers ($f_{\alpha\sigma}=1$ or 0).
The density for spin $\sigma$ is $n_\sigma(\bfr)=\rho_{\sigma}(\bfr,\bfr)$.

A little-known aspect of the work of Perdew and
Zunger~\cite{Perdew:1981/PRB/5048} is their guess that
Eqs.~(\ref{exden})--(\ref{matrix}) apply, with fractional occupations
$0\le f_{\alpha\sigma}\le 1$, even to an open \textit{subsystem}
(with average electron number $N=\sum_{\alpha\sigma}f_{\alpha\sigma}$)
of a closed system. We confirm this guess numerically now and
analytically later. Numerically, we find~\cite{Vydrov:2007/JCP/154109}
that the total Hartree-Fock energy computed self-consistently
by Eqs.~(\ref{exden})--(\ref{matrix}) for a molecule A$_2^{+}$
at $R \rightarrow \infty$ is exactly twice that of A$^{+0.5}$.
(The difference between Hartree-Fock and exact exchange-only DFT
energies is unimportant on the scale of the effects we study).

\begin{figure}
\includegraphics[width=\columnwidth]{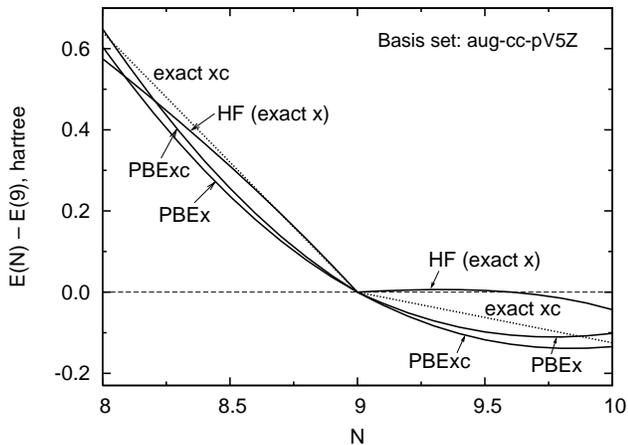}
\caption{\label{fig:1}
Total energies of the F atom as functions of the average electron
number $N$ in the PBE approximation (with and without correlation)
and in Hartree-Fock theory. The dotted lines represent the exact-xc
result based on the experimental ionization potential and electron
affinity of the F atom~\cite{Perdew:1982/PRL/1691}.}
\end{figure}

Consider the total energy of an F atom, treated as an open
system, as a function of average electron number $N$. We vary $N$ by
changing the population of the highest partly-occupied $2p$-orbital
and compute the energy self-consistently using the \textsc{gaussian}
program~\cite{GDV-E05}. Fig.~\ref{fig:1} shows that the total energy
with exact exchange-correlation varies linearly between adjacent
integers~\cite{Perdew:1982/PRL/1691,Perdew:1985/Dreizler},
but the PBEx and PBExc energies curve upward strongly, while the
HF energy curves downward. Note also the accuracy of PBExc
energy differences for integer values of $N$ around 9 and their
inaccuracy for non-integer values. LSD~\cite{Perdew:1985/Dreizler},
TPSS~\cite{Vydrov:2007/JCP/154109}, and other
functionals~\cite{Mori-Sanchez:2006-2007} behave similarly.

In Fig.~\ref{fig:1}, the Hartree-Fock approximation (exact exchange
without correlation) shows substantial midpoint error for non-integer
electron numbers. However, this error is positive and thus not
very harmful since energy minimization forces integer electron
numbers onto separated open subsystems, e.g., a symmetry-broken
F$\cdots$F$^{+}$ as the dissociation limit of F$_2^{+}$. But
F$^{+0.5}\cdots$F$^{+0.5}$ is a harmfully too-deep minimum for
the semi-local density functionals, since their midpoint error
is negative.

\begin{figure}
\includegraphics[width=\columnwidth]{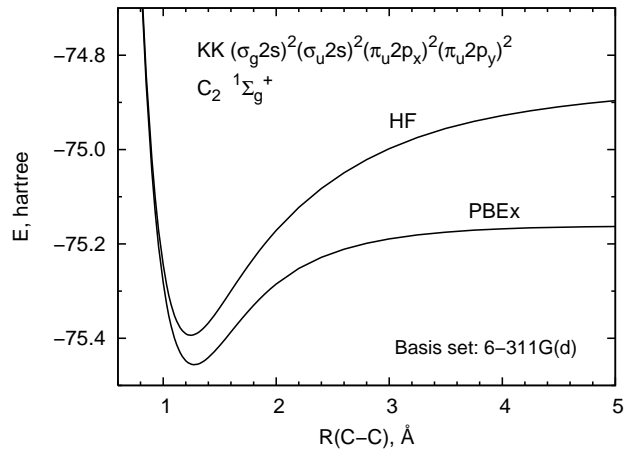}
\caption{\label{fig:2}
Spin-restricted exchange-only dissociation curves for
the C$_2$ molecule in the symmetric singlet electron configuration.
This configuration is the lowest-energy state in the PBEx
approximation near the equilibrium internuclear distance
but not at larger $R$, and not at any $R$ in Hartree-Fock.}
\end{figure}

The failure of semi-local exchange can occur even in open systems
with integer average electron number. In semi-local DFT, the
correct dissociation limit for neutral homonuclear diatomics built
up from open-shell atoms (e.g., H$_2$) is achieved by spin symmetry
breaking. If, however, spin symmetry is imposed, then the semi-local
exchange energy in the $R\to\infty$ limit is much more negative than
the exact exchange energy~\cite{Handy:2001/MP/403}. Typically,
the separated atoms have half-integer numbers of electrons
of each spin (e.g., $N_\uparrow$=$N_\downarrow$=$\frac{1}{2}$
on each spin-unpolarized H atom). But the C$_2$ molecule in the
unusual singlet configuration $KK (\sigma_g 2s)^2 (\sigma_u 2s)^2
(\pi_u 2p_x)^2 (\pi_u 2p_y)^2$ dissociates to two neutral C
atoms, each with $N_\uparrow$=$N_\downarrow$=3 but fractional
occupations $f_{2p_x\uparrow}$=$f_{2p_x\downarrow}$=$
f_{2p_y\uparrow}$=$f_{2p_y\downarrow}$=$\frac{1}{2}$.
Fig.~\ref{fig:2} shows that, again, the spin-restricted semi-local
exchange energy in the $R\to\infty$ limit is far more negative than
that of spin-restricted Hartree-Fock theory.

In a transition-state complex of a chemical reaction, residual
fluctuation of electrons among its weakly-bonded fragments raises
the total energy via increased Coulomb repulsion, but semi-local
exchange approximations miss this effect and predict reaction
barriers that are too low.

When we apply single-configuration Hartree-Fock theory to a closed
system, fractional occupation numbers on a separated open subsystem
can \textit{only} represent fluctuation of electrons among separated
subsystems. That is the case we address here. In a different case,
fractional occupation on a closed system can represent fluctuation
of electrons among degenerate orbitals of that system.

We will now prove that Eqs.~(\ref{exden})--(\ref{matrix})
apply to an open system of fluctuating electron number and then
show how they imply the behavior found in Figs.~\ref{fig:1}
and~\ref{fig:2}.  Let S be a fully separated open system and R
its reservoir, i.e., the only other open system with which S is
free to exchange electrons. Consider now the closed system S+R of
integer electron number, which we will describe by a single Slater
determinant.  If the wavefunction of S+R were fully correlated,
the ensemble describing S as a subsystem of S+R would also
be~\cite{Perdew:1985/Dreizler}, but since the former is only a
single energy-minimized Slater determinant, the latter consists
of Slater determinants and their probabilities that need not be
energy-minimized.  When we form S+R, each orthonormal orbital on S
remains unchanged if there is no orbital of the same energy from
R, or hybridizes (forms a linear combination) with orbitals of
the same energy from R, resulting in a band whose width tends
to zero in the infinite-separation limit (i.e., a group of
nearly degenerate molecular orbitals). If this band is filled,
hybridization is just a unitary transformation of the occupied
orbitals.  The Hartree-Fock exchange energy of S+R, as predicted
by Eqs.~(\ref{exden})--(\ref{matrix}), is a sum of a pure S term
(arising when $\bfr$ and $\bfr'$ are on S) and a pure R term (arising
when $\bfr$ and $\bfr'$ are on R), because when $\bfr$ is in S and
$\bfr'$ in R the Coulomb interaction $1/|\bfr-\bfr'|$ vanishes.
Thus, the S term is given by Eqs.~(\ref{exden})--(\ref{matrix})
in which $\psi_{\alpha\sigma}$ are the orthonormal orbitals of
S with $f_{\alpha\sigma}=1$ if the corresponding band is filled
and $0<f_{\alpha\sigma}<1$ if it is partly filled (holds more
nearly-degenerate orbitals than electrons), as it can be if the
Fermi level lies within the band. This concludes the proof.

Now we can use the ensemble describing S to explain why
the Hartree-Fock energy as a function of $N$ curves downward.
Let $E_i^\mathrm{HF}(\mbox{S};\mbox{S+R})$ be the Hartree-Fock
energy of the $N_i$-electron pure state $i$ of system S,
evaluated from orbitals for S formed by truncating and
renormalizing the ground-state molecular orbitals of S+R,
and let $p_i$ be the probability to find state $i$ in the ground
state of S+R, with $\sum_i p_i = 1$ and $\sum_i p_i N_i=N$.
Then $E^\mathrm{HF}(\mbox{S};\mbox{S+R})
 =\sum_i p_i E_i^\mathrm{HF}(\mbox{S};\mbox{S+R}) \ge
\sum_i p_i E_i^\mathrm{HF}(\mbox{S};\mbox{S})$.
A familiar and instructive example~\cite{Book/Levine:1991}
is the H atom in a spin-restricted stretched H$_2$ molecule,
where the states $i$ are the neutral atoms of each spin
($p_i=0.25$, $N_i=1$), cation ($p_i=0.25$, $N_i=0$),
and anion ($p_i=0.25$, $N_i=2$). Another
example~\cite{Balawender:2005/JCP/124102} is Fig.~\ref{fig:1},
where the states $i$ are those for $N_i=J$ and $J-1$ electrons,
with $J$ an integer. 

The reason why semi-local functionals predict too-negative
energies for systems with fractional occupations also becomes
clear. By the orthonormality of the orbitals of a subsystem,
Eqs.~(\ref{xhole}) and~(\ref{matrix}) imply the sum rule 
\begin{equation}\label{sumr}
\int d\bfr' \ n_\mathrm{x}(\bfr,\bfr')
  = -\sum_\sigma\sum_\alpha \frac{f_{\alpha\sigma}
 n_{\alpha\sigma}(\bfr)}{n(\bfr)},
\end{equation}
where $n_{\alpha\sigma}(\bfr)
 = f_{\alpha\sigma}|\psi_{\alpha\sigma}(\bfr)|^2$. 
Eq.~(\ref{sumr}) for noninteger $f_{\alpha\sigma}$
was presented without proof in Ref.~\onlinecite{Perdew:1981/PRB/5048}.
Adding and subtracting $-1 = -\sum_{\alpha\sigma}
n_{\alpha\sigma}(\bfr)/n(\bfr)$ to Eq.~(\ref{sumr}), we find
\begin{equation}\label{sumr2}
\int d\bfr'\, n_\mathrm{x}(\bfr,\bfr')
 = -1 + \sum_{\alpha\sigma} f_{\alpha\sigma}
 (1 - f_{\alpha\sigma}) \frac{|\psi_{\alpha\sigma}(\bfr)|^2}{n(\bfr)}.
\end{equation}
The close similarity between the sum rule integral [left-hand
side of Eq.~(\ref{sumr2})] and the exchange-energy-per-electron
integral [right-hand side of Eq.~(\ref{exden})] for
integer $f_{\alpha\sigma}$ was pointed out by Gunnarsson and
Lundqvist~\cite{Gunnarsson:1976/PRB/4274}, and this similarity
persists for noninteger values of $f_{\alpha\sigma}$.  When
all the occupation numbers are 1 or 0, the right-hand side of
Eq.~(\ref{sumr2}) becomes $-1$, which is also the sum rule implicitly
assumed by LSD, PBE, and TPSS~\cite{Constantin:2006/PRB/205104}.
But, when some occupation numbers are between 1 and 0, the right-hand
side of Eq.~(\ref{sumr2}) will fall between $-1$ and 0, meaning
that part of the exact exchange hole around an electron in an
open system is located in its distant reservoir. In this case a
semi-local exchange approximation $\epsilon_\mathrm{x}^\mathrm{sl}$
will tend to be more negative than the exact exchange energy per
electron $\epsilon_\mathrm{x}^\mathrm{ex}(\bfr)$, as illustrated in
Figs.~\ref{fig:1} and~\ref{fig:2}.  The total or system-averaged
$E_\mathrm{x}^\mathrm{sl}$ also violates the sum rule on the
system-averaged~\cite{Tao:2003/JCP/6457} x-hole
\begin{equation}
 \frac{1}{N} \int d\bfr' d\bfr\; n(\bfr) n_\mathrm{x}(\bfr,\bfr')
 = -1 + \frac{1}{N}\sum_{\alpha\sigma} f_{\alpha\sigma}
 (1 - f_{\alpha\sigma}).  \label{sumr3}
\end{equation}
The Hartree-Fock mean square fluctuation of electron number,
$\sum_{\alpha,\sigma} f_{\alpha\sigma} (1-f_{\alpha\sigma})$,
only vanishes when all occupations are 0 or 1.

A sum rule for the exact xc-hole density $n_\mathrm{xc}(\bfr,\bfr')$
is also known~\cite{Perdew:1985/Dreizler}. Its integral equals $-1$
only when the electron number does not fluctuate and otherwise falls
between $-1$ and 0. Ref.~\onlinecite{Perdew:1985/Dreizler}
presents a coupling-constant integration for $E_\mathrm{xc}$
and $n_\mathrm{xc}$.
But the integrand for $E_\mathrm{xc}$ at zero coupling
strength is not really the exact exchange-only energy because of an
exact-degeneracy static correlation. When the electron number on the
infinitely-separated open system S fluctuates at the Hartree-Fock
level, occupied and unoccupied orbitals (with the same spin) of
closed system S+R are degenerate. Degenerate perturbation theory is
needed to find the correlation energy, which is of the same order
as the exchange energy even in the weak-coupling or high-density
limits. Exact-degeneracy correlation and normal correlation
shift the downward-curved Hartree-Fock energy of Fig.~\ref{fig:1}
into the straight-line correlated exact energy. Note also from
Fig.~\ref{fig:1} that semi-local approximations for $E_\mathrm{xc}$
overestimate the strength of exact-degeneracy correlation (which
they introduce via $E_\mathrm{x}^\mathrm{sl}$).

Semi-local functionals are often combined with a Hubbard $U$
parameter (DFT+$U$).  A simple case occurs when only one localized
orbital has non-integer occupation $f$, and the method adds to the
semi-local energy a positive term $U f(1-f)$. The close connection
between DFT+$U$ and self-interaction correction has been
argued~\cite{Cococcioni:2005/PRB/035105}.
We note that $U$ does not represent ``strong
correlation'' (as sometimes asserted), because the $U$ needed to
reach the Hartree-Fock energy is greater than that needed to reach
the exact correlated energy. $U$ favors the less fluctuating
configuration by penalizing the more fluctuating one.

Our Fig.~\ref{fig:1} and our analysis explaining it show that some
region-dependent fraction (between 0 and 1) of exact exchange is
needed to correct semi-local exchange-correlation approximations.
Such a mixing of the downward-curving exact exchange with
the upward-curving semi-local exchange and semi-local correlation
can produce the needed straight line.
This motivates a local hybrid (lh) functional
\begin{equation}
\epsilon_\mathrm{xc}^\mathrm{lh}(\bfr)
  = \epsilon_\mathrm{x}^\mathrm{ex} +
 [1 - a(\bfr)](\epsilon_\mathrm{x}^\mathrm{sl}
 - \epsilon_\mathrm{x}^\mathrm{ex})
 + \epsilon_\mathrm{c}^\mathrm{sl},   \label{lh}
\end{equation}
where $0 \le a(\bfr) \le 1$ and sl=TPSS. Eq.~(\ref{lh})
was presented in Ref.~\onlinecite{Cruz:1998/JPCA/4911}
without a form for $a(\bfr)$. Forms were
proposed in Ref.~\onlinecite{Perdew:2001/1} and in
Ref.~\onlinecite{Jaramillo:2003/JCP/1068} (where the term ``local
hybrid" was coined), but did not achieve useful accuracy for
equilibrium properties~\cite{Jaramillo:2003/JCP/1068}. The choice
$a(\bfr) = 1$ satisfies nearly all universal constraints but
misses the delicate and helpful error cancellation between
semi-local exchange and semi-local correlation that typically
occurs (because the xc-hole is deeper and more short-ranged than
the x-hole) in normal regions where the density is not too high,
too strongly-varying, too one-electron-like, or too fluctuating
in a spin-polarized region at the Hartree-Fock level (i.e.,
$\epsilon_\mathrm{x}^\mathrm{ex}/\epsilon_\mathrm{x}^\mathrm{sl}\ll
1$).  So we will take $a(\bfr)$ to be small or 0 in a normal
region, and to tend toward 1 to the extent that any condition
of normality is violated. Then the second and third terms on
the right of Eq.~(\ref{lh}) represent~\cite{Handy:2001/MP/403}
near-degeneracy static and dynamic correlation, respectively. The
dominance of exact exchange [$a(\bfr)\to 1$] in the high-density
limit means that Eq.~(\ref{lh}) has 100\% exact exchange plus fully
nonlocal correlation.

A natural generalization of the global hybrid of Eq.~(\ref{global}),
Eq.~(\ref{lh}) can satisfy~\cite{Perdew:2001/1,HGGA:xxx} many more exact
constraints while achieving greater accuracy. We have developed
and are testing~\cite{HGGA:xxx} such a local hybrid hyper-GGA,
and the results will be reported later. One or more universal
empirical parameters are needed, as in Eq.~(\ref{global}), since
the universal constraints are already satisfied by $a(\bfr) = 1$.
As usual, symmetry must be allowed to break.  Spin-symmetry
breaking~\cite{Perdew:1995/PRA/4531},
like real correlation, lowers the energy by
suppressing Hartree-Fock-level fluctuation of electron number.
Semi-local functionals mimic this suppression.  Where real
correlation cannot fully suppress fluctuation, as in open systems
of non-integer average electron number, semi-local functionals
overcorrelate and need a large positive correction, i.e., a large
$a(\bfr)$.

The Perdew-Zunger self-interaction
correction~\cite{Perdew:1981/PRB/5048} to semi-local functionals
works in much the same way to raise the energy of a system with
fractional occupation~\cite{Ruzsinszky:2006-2007}, satisfying the sum
rule of Eq.~(\ref{sumr2}). However, it loses the error cancellation
between semi-local exchange and semi-local correlation in normal
regions, and so is inaccurate for molecules near equilibrium
geometries~\cite{Vydrov:2004/JCP/8187}.

In summary, striking and diverse failures of semi-local functionals
arise because they assign too low an energy to configurations
where the electron number in a spin-polarized region fluctuates
too strongly (i.e., where $\epsilon_\mathrm{x}^\mathrm{ex}
/\epsilon_\mathrm{x}^\mathrm{sl}\ll 1)$ at the Hartree-Fock
level. These errors should be corrected by \textit{local} mixing
of a variable fraction of exact exchange.

This work was supported by NSF Grants DMR-0501588 (J.P.P.)
and CHE-0457030 (G.E.S. and O.A.V.), OTKA Grant PD-050014 (A.R.),
OTKA-NSF (G.I.C.), the Welch Foundation (G.E.S.),
and by DOE Grant DE-FG02-05ER46203 (J.T.)


\end{document}